\documentstyle[multicol,eqsecnum,prd,aps,graphicx]{revtex}    
\begin{document}
\draft
\title{
 A texture of neutrino mass matrix in view 
        of recent neutrino experimental results
}       
\author{\bf Ambar Ghosal\thanks{E-mail address : ambar@anp.saha.
ernet.in}  and Debasish Majumdar\thanks{E-mail address: debasish@
theory.saha.ernet.in}} 
\address{ 
        Saha Institute of Nuclear Physics,
        1/AF Bidhannagar, Kolkata 700 064, India 
}
\date{\today}
\maketitle
\begin{abstract} 
In view of recent neutrino experimental
results such as SNO, Super-Kamiokande (SK),
CHOOZ and neutrinoless double beta decay $(\beta\beta_{0\nu})$
,we consider a  texture of neutrino mass
matrix which contains three parameters in order to explain
those neutrino experimental results.
We have first fitted parameters in a model
independent way with solar and atmospheric neutrino
mass squared differences and solar neutrino mixing angle which
satisfy LMA solution.
The maximal value of atmospheric neutrino
mixing angle comes out naturally in the present texture.
Most interestingly, fitted parameters of the  neutrino mass
matrix considered here 
also marginally satisfy recent limit on effective Majorana neutrino mass
obtained from neutrinoless double beta decay experiment.
We further demonstrate an
explicit model which gives rise to the texture investigated
by considering an $SU(2)_L\times U(1)_Y$
gauge group with two extra real scalar singlets and discrete
$Z_2\times Z_3$ symmetry. Majorana neutrino
masses are generated through
higher dimensional operators at the scale $M$.
We have estimated the scales at which singlets get VEV's and M by
comparing with the best fitted results obtained in the present
work.
\vspace{1pc}
\end{abstract}
\maketitle
\maketitle
\begin{multicols}{2}
\narrowtext
\section{
 Introduction}
A recent global analysis
\cite{Bahcall}
including SNO experimental results containing
neutral current data of solar neutrino flux,
day night effect and higher statistics data of
charged current 
neutrino electron scattering rate,
\cite{sno}
has disfavoured the well known conjecture of
bimaximal neutrino mixing
\cite {bimax,ag1}
by considering the solution
of solar
neutrino problem through two flavour neutrino oscillation
scenario.
It has been shown 
that the best fit global oscillation parameters
with all solar neutrino experimental data strongly
in favour
of 
the large angle
MSW oscillation solution (LMA) of solar neutrino
deficit and the best fit result comes out as 
$\Delta m^2_\odot$ = $5.0\times {10}^{-5}$ $e{\rm V^2}$,
$\tan^2\theta_\odot$ = $4.2\times {10}^{-1}$
with the value of
$\chi^2_{\rm min}$ = 45.5 and g.o.f. = 49\%.
Although, the atmospheric neutrino oscillation mixing
angle $\theta_{\rm atm}$
is maximal as observed by Super-Kamiokande(SK)
\cite{SK,oba}
however, the LMA solution for the solar neutrino
oscillation is
best fitted with a considerably lower value
of $\theta_\odot$.
If we identify the $\theta_\odot$ as $\theta_{12}$ and
$\theta_{\rm atm}$ as $\theta_{23}$ then 
the CHOOZ
\cite{chooz}
experiment has also constrained the third mixing angle
$\theta_{13}< 13^o$.
Furthermore, recent result from neutrinoless double beta decay
$(\beta\beta_{0\nu})$ experiment \cite{2beta} 
has reported the bound on
the effective Majorana neutrino mass ( by considering uncertnity 
of the nuclear matrix elements upto $\pm 50\%$ and the contribution 
to this process due to particles other than Majorana neutrino is 
negligible \cite{nonma}) as  
$$
\langle m\rangle \,\,\,
= \,\,\,(\,\,0.05\,\, - \,\,0.84\,\,)
\,\,\,{\rm eV}
\,\,\,\,{\rm at}\,\,\, 95\% \,\,\,{\rm c.l.}
\eqno(1.1)
$$
Although, there are several number of literature investigating the 
implications of the above experimental result, 
however, the claim is still  
controversial  \cite{reply}.  
In the present work, we go optimistically with the result and it is 
important to note that  
our analyses crucially depends on the future results of MOON, EXO , 
1 ton and 10 ton GENIUS double beta decay experiments. If the lower 
limit of the Majorana neutrino mass goes below the value presented 
in Eq.(1), the texture and the model considered here 
will be ruled out provided there must not be significant changes   
in the present solar and atmospheric neutrino experimental results.
Two parameter texture of neutrino
mass matrix
\cite{ag1,two}
which naturally gives bi-maximal neutrino
mixing is disfavoured by the recent SNO experimental results.
It needs more parameters in the neutrino mass matrix in order
to explain the present neutrino experimental results
\cite{yasue}. 
Moreover, in order to satisfy the limit
on effective Majorana neutrino mass the $\nu_e\nu_e$ element
of the neutrino mass matrix should be non-zero.
All these results need further modification of two parameter
neutrino mass matrix.
\vskip .1in
\noindent
In the present work, we consider a three parameter
texture of
neutrino mass matrix which can accommodate the present
experimental results. 
These three parameters are fixed 
by defining a function $\chi^2_p$ (see later) as 
the sum of squares of the differences between the calculated values 
of neutrino oscillation parameters (with the texture considered) and 
the best fitted values of the same (obtained from different analyses) 
and then minimising this function. 
We then propose an explicit
model based on an $SU(2)_L\times U(1)_Y$ gauge group with
two additional singlet real scalar fields and discrete
$Z_2\times Z_3$ symmetry. Neutrino mass is generated in our
model through higher dimensional terms, the scale of which
is fixed through best fit result.

\section {A Texture}
Keeping in mind the charged lepton mass matrix is diagonal
, consider the following texture of the Majorana
neutrino mass
matrix
$$
M_\nu = \lambda\left (
\begin{array}{ccc}
b&a&a\\
a&1&1\\
a&1&1
\end{array}\right ) \
\eqno(2.1)
$$
where $\lambda$, $a$ and $b$ are all real. It is to be noted that 
a more general form of the above neutrino mass matrix is presented 
in Ref.[11] from which under certain conditions of 
model parameters the above 
form can be obtained. Next, 
particularly, due to the choice of $a$ and $b$ parameters
as real, the above neutrino mass matrix gives no CP
violation effects in the leptonic sector. Furthermore, the
$\nu_e\nu_e$   element of the $M_\nu$ is non-zero hence,
it should give rise to $\beta\beta_{0\nu}$ decay. We will
estimate the constraint on the $\nu_e\nu_e$ element from 
$\beta\beta_{0\nu}$ decay after fitting the solar and
atmospheric neutrino experimental results. Moreover, the
above neutrino mass matrix can be generated either by
radiative way or by non-renormalisable operators.
\par
\noindent
Diagonalising the above neutrino mass matrix $M_\nu$
by an orthogonal transformation as
$$
O^T M_\nu O = {\rm Diag}( m_1, m_2, m_3)
\eqno(2.2)
$$
where
\vspace*{3mm}
\noindent
$O = $
$$\pmatrix{
c_{31}c_{12} & c_{31}s_{12}& s_{31} \cr
&&\cr
-c_{23}s_{12} &c_{23}c_{12}&s_{23} c_{31}\cr
 - s_{23}s_{31}c_{12} &
- s_{23}s_{31}s_{12} &
\cr
&&\cr
s_{23}s_{12}&-s_{23}c_{12}&c_{23} c_{31}\cr
- c_{23}s_{31}c_{12} &
- c_{23}s_{31}s_{12} &
}\,\,,
\eqno(2.3)
$$
we obtain the following mixing angles
$$
\theta_{23} = -\pi/4, \theta_{31} = 0
\eqno(2.4a)
$$
$$
\tan^2\theta_{12} = \frac{\lambda b - m_1}{m_2 - \lambda b}
\eqno(2.4b)
$$
and the eigenvalues are
$$
m_1 = {\frac{\lambda}{2}}\left \{ (2+b) +
\sqrt{{(2-b)}^2 + 8a^2}\right \}
$$
$$
m_2 = {\frac{\lambda}{2}}\left \{ (2+b) -
\sqrt{{(2-b)}^2 + 8a^2}\right \}
$$
$$
m_3 = 0
\eqno(2.5)
$$
\noindent
Furthermore, in terms of
these eigenvalues the mixing matrix
$O$ can be written as
$$
O = 
 \left (
\begin{array}{ccc}
c_{12} & s_{12}& 0 \\
-\frac{s_{12}}{\sqrt{2}}&
\frac{c_{12}}{\sqrt{2}}&
-\frac{1}{\sqrt{2}} \\
-\frac{s_{12}}{\sqrt{2}}&
\frac{c_{12}}{\sqrt{2}}&
\frac{1}{\sqrt{2}} 
\end{array} \right )
$$
$$=
\left(
\begin{array}{ccc}
{\sqrt{\frac{m_2 - \lambda b}{ m_2 - m_1}}}
&
{\sqrt{\frac{\lambda b - m_1}{ m_2 - m_1}}}
&
0
\\
-{\frac{1}{\sqrt{2}}}
{\sqrt{\frac{\lambda b - m_1}{ m_2 - m_1}}}
&
{\frac{1}{\sqrt{2}}}
{\sqrt{\frac{m_2 - \lambda b}{ m_2 - m_1}}}
&
-{\frac{1}{\sqrt 2}} \\

-{\frac{1}{\sqrt{2}}}
{\sqrt{\frac{\lambda b - m_1}{ m_2 - m_1}}}
&
{\frac{1}{\sqrt{2}}}
{\sqrt{\frac{m_2 - \lambda b}{ m_2 - m_1}}}
&
{\frac{1}{\sqrt 2}} 
\end{array} \right) 
\eqno(2.6)
$$
We set the solar and
atmospheric neutrino mass squared
differences as
$$
\Delta m_{sol}^2 = \Delta m_{21}^2 = m_1^2 - m_2^2
$$
$$
= \lambda^2 (2+b)\sqrt{{(2-b)}^2 + 8a^2}
\eqno(2.7a)
$$
and
$$
\Delta m_{\rm atm}^2 = \Delta m_{23}^2 = m_2^2 - m_3^2
= m_2^2
\eqno(2.7b)
$$
\noindent
The best fit values of oscillation
parameters from solar neutrino experiment
and atmospheric neutrino
experiments are used to obtain the values of
$a$, $b$ and $\lambda$.  For this purpose, we consider 
$\Delta m_{12}^2$, the 
difference of the square
of mass eigenstates $m_1$ and $m_2$ and
the mixing angle $\theta_{12}$
are responsible for solar neutrino oscillation
and  $\Delta m_{23}^2$  and
$\theta_{23}$ are responsible for oscillation
of atmospheric neutrinos.
A recent global analysis by Bahcall et al \cite{Bahcall} 
of the solar neutrino data
from all solar neutrino experiments namely Chlorine, Gallium, 
Super-Kamiokande, SNO charged current including the recently 
published SNO neutral current 
data, shows that large mixing angle or
LMA solution is most
favoured for solar neutrino oscillation. According to this
analysis $\Delta m_\odot^2 (\equiv \Delta m_{12}^2$
for our model) $= 5 \times
10^{-5}$ and tan$^2\theta_\odot
(\equiv$ tan$^2\theta_{12}$ for our model)
$= 0.42$. From the analysis of
atmospheric neutrino oscillation data
\cite{SK} we have the best fit values of $\Delta m_{\rm atm}^2 
(\equiv \Delta m_{23}^2$
for our model) $= 3.1 \times 10^{-3}$ eV$^2$ and $\theta_{\rm atm}$
maximal. This value of $\theta_{\rm atm}$
has already been obtained in our model
for $\theta_{23}$. Thus treating $a$, $b$ and $\lambda$ as 
parameters we can obtain different values of    
$\Delta m_{12}^2$, $\Delta m_{23}^2$ and tan$^2\theta_{12}$ and 
compare them with best fit values of those quantities namely 
$\Delta m_\odot^2$, $\Delta m_{\rm atm}^2$ and tan$^2\theta_\odot$
obtained from solar and atmospheric neutrino analysis of 
data (discussed above)
to fix $a$, $b$ and $\lambda$. To this end we define a function
$$
\chi^2_p = (\Delta m_{12}^2 - \Delta m^2_\odot)^2 
+ (\Delta m_{23}^2 - \Delta m^2_{\rm atm})^2  $$ $$ 
 +  (\tan^2\theta_{12} - \tan^2\theta_\odot)^2\,\, .
\eqno(2.8)
$$
The function $\chi^2_p$ as defined above is calculated for 
a wide range of values of 
$a$, $b$ and $\lambda$ and the minimum of the function is obtained.
The corresponding values of $a$, $b$ and $\lambda$ are given below.
$$
{\rm Minimum\,\,} \chi^2_p = 1.\,\,4 \times 10^{-8}
$$
$$
\,\,\,\,\,\,\,\,\,\,\,\,\,\,\,\,\,\,\, a \,\, = 0.\,\,0142 
$$
$$
\,\,\,\,\,\,\,\,\,\,\,\,\,\, b \,\,\,\, =  2.\,\,018
$$
$$
\,\,\,\,\,\,\,\,\,\,\,\,\,\,\,\,\,\,\,\,\,
\,\, \lambda \,\,\,\,= 0.\,\,028\,\,\, {\rm eV}
\eqno(2.9)
$$
\noindent
$\Delta m_{12}^2$, $\Delta m_{23}^2$ and $\tan^2\theta_{12}$ obtained 
from the above values of $a$, $b$ and $\lambda$ and their comparison 
with the best fit values for 
$\Delta m^2_\odot$, $\Delta m^2_{\rm atm}$ and $\tan^2\theta_{\odot}$  
obtained from recent analysis of the solar and atmospheric neutrino data
are shown in Table 1.
\noindent
In order to find out the range
of values of $a$ and $b$ that satisfy the
3$\sigma$ limits of $\Delta m_\odot^2$
and $\tan^2\theta_\odot$ for LMA solution
(Eq. (2.1) and (2.3) of \cite{Bahcall})
of combined analysis of recent
solar neutrino data (including SNO neutral current data),
we have fixed the value
of $\lambda$ at 0.028 (see Eq. (2.9)) and varied $a$
and $b$ so that
$\Delta m^2_{12}$ and $\tan^2\theta_{12}$ satisfy the allowed
LMA solution range mentioned above.
This range as given in Ref. \cite{Bahcall}
is $2.3 \times 10^{-5} < \Delta m_\odot^2 < 3.7 \times 10^{-4}$ and 
$0.24 < \tan^2 \theta_\odot < 0.89$.  
In doing this $\Delta m_{23}^2$ remains fixed at 
$3.1 \times 10^{-3}$ -- the value for atmospheric neutrino solution.
The allowed region in parameter
space of $a$ and $b$ is shown in Fig. 1.
We find that the allowed parameter space is very sensitive on
$\lambda$, and there is not much freedom to vary $\lambda$
within
a wide range. 
Next, we consider the bound on $\nu_e\nu_e$ matrix element of
$M_\nu$ from $\beta\beta_{0\nu}$ decay experiment.
In the present work, after fitting all three prameters with
solar and atmospheric neutrino experimental results, we find that
 the value of effective neutrino mass comes out as 
$$
\langle m_\nu\rangle
\,\,\,= \,\,\,\lambda \,\,b
\,\,\,= \,\,\,0.\,\,05 \,\,\, {\rm eV}\,\, 
\eqno(2.10)
$$
\noindent
which is marginally at the lower end of the experimental
value. Such value may be accidental or may have some deeper
meaning however, it is quite interesting to note that the
testability of the present texture crucially lies on the
future result of the $\beta\beta_{0\nu}$ experiment.
\section{ A Model}
We consider an $SU(2)_L\times U(1)_Y$ model with
two additional singlet real scalar fields and discrete
$Z_2\times Z_3$ symmetry. The representation content of
the leptonic and scalar fields is given in Table 2.
Apart from the Standard Model (SM) Higgs doublet , the
extra singlets considered in the present model give rise
to three parameters in the neutrino mass matrix. The charged
lepton masses generated in the present model is similar
to those in SM. To make the charged lepton mass matrix flavour
diagonal we consider a reflection symmetry on the
lepton- Higgs Yukawa
coupling $f_{ij}$ ($i,j$ = 1,2,3 flavour indices ) as
$$
f_{ij}\leftrightarrow f_{ji}, i\neq j .
\eqno(3.1)
$$
\noindent
We consider soft discrete symmetry breaking  terms
( Dim$\leq$ 3) in the scalar potential of the model,
and , hence, none of the VEV is zero upon
minimisation of the scalar potential \cite{ag1,ag}.
In the present model, Majorana neutrino masses are
obtained through higher dimensional terms due to
explicit violation of lepton number. The most general
lepton-scalar Yukawa interaction in the present model
generating Majorana neutrino masses is given by
$$
L_Y^\nu =
\frac{l_{1L} l_{1L}\phi\phi\eta_1}{M^2} 
+
\frac{l_{1L} l_{2L}\phi\phi\eta_2}{M^2}
+
\frac{l_{1L} l_{3L}\phi\phi\eta_2}{M^2}
+
$$
$$\frac{l_{2L} l_{2L}\phi\phi}{M}
+
\frac{l_{2L} l_{3L}\phi\phi}{M}
+
\frac{l_{3L} l_{3L}\phi\phi}{M} 
\eqno(3.2)
$$
\noindent
and the charged lepton masses are generated through
the following interaction
$$
L_Y^E = f_{11}{\bar l_{1L}} e_R \phi
+
f_{22}{\bar l_{2L}} \mu_R \phi
+
f_{33}{\bar l_{3L}} \tau_R \phi
+ h.c.
\eqno(3.3)
$$
\noindent
All other terms in Eq. (3.2) are prohibited due to 
discrete $Z_2\times Z_3$ symmetry and reflection symmetry
mentioned in Eq. (3.1). Substituting VEV's of the scalar
fields in Eqs. (3.3) and (3.2) we obtain respectively
\noindent
$$
M_E = \pmatrix{f_{11}\langle\phi\rangle&0&0\cr
               0& f_{22}\langle\phi\rangle&0\cr
               0&0& f_{33}\langle\phi\rangle}
\eqno(3.4)
$$
$$
M_\nu = \lambda\left(
\begin{array}{ccc}
b&a&a\\
a&1&1\\
a&1&1
\end{array}
\right) \
\eqno(3.5)
$$
with
$$
\lambda = \frac{{\langle\phi\rangle}^2}{M},
b = \frac{\langle\eta_1\rangle}{M}, 
a = \frac{\langle\eta_2\rangle}{M}.
\eqno(3.6)
$$
\noindent
From our best fitted results given in Eq. (2.9)
which satisfy the
LMA solution of Solar
neutrino solution  and atmospheric neutrino
experimental results, we obtain the
parameters $M$, $\langle\eta_1\rangle$,
$\langle\eta_2\rangle$
as $M$ = $3.5\times 10^{14}$ GeV,
$\langle\eta_1\rangle$ =
$7\times 10^{14}$ GeV,
$\langle\eta_2\rangle$ =
$4.97\times 10^{12}$ GeV
with $\langle\phi\rangle$ = 100 GeV.
It is to be noted that although the value of
$\langle\eta_1\rangle$
comes out to be greater than the effective scale $M$,
the effective coupling is always less
than unity due to the factor
$\lambda$ and the
effective coupling is always within the perturbative
limit.
Apart from the
electroweak scale, the present model contains three
other mass scales two of which are very near
to SUSY unification scale and the third is little
lower. One of the singlet gets VEV at little 
above the effective
scale $M$ of the theory. It has some analogy with
the singlets getting VEV's between SUSY unification
scale and Planck scale in supersymmetric theory.

\section{Conclusion}
In view of the results from solar neutrino experiments including
recent SNO neutral current experiment, results from atmospheric
neutrino experiment 
and CHOOZ experimental results we consider a texture of neutrino
mass matrix which contains three parameters. We have considered
no CP violation effects by choosing these parameters as real.
The atmospheric neutrino mixing angle $\theta_{23}$
comes out to be maximal
and $\theta_{13}$ = 0. The later satisfies CHOOZ experimental results.
We have fixed the three parameters of the texture considered by fitting
the mass squared differences corresponding to solar and atmospheric 
neutrinos and the mixing angles corresponding to solar neutrinos 
with the best fit values of those quantities (LMA solution for solar 
neutrinos).
We also find that the best fitted parameters can 
accommodate very marginally the
bound on effective neutrino mass from neutrinoless double beta
decay experiment and thus the testability of the present
texture lies crucially on the future result of
$\beta\beta_{0\nu}$ experiment.
We also demonstrate an explicit model based on an
$SU(2)_L\times U(1)_Y$ gauge group with extended Higgs sector
and discrete symmetry that gives rise to the
texture of neutrino mass matrix investigated along with diagonal
charged lepton mass matrix. Neutrino masses are generated through
higher dimensional operators at the scale $M$.
Comparing with the best fitted values
obtained, we
estimate the scale of the VEV's of the neutral singlet
scalars as $\langle\eta_1\rangle$ = $7\times 10^{14}$ GeV,
$\langle\eta_2\rangle$ = $4.97\times 10^{12}$ GeV
and the scale $M$ = 3.5$\times 10^{14}$ GeV.
We will further study how such texture
can be  realized within GUT, SUSYGUT scenarios. 

\vspace*{3mm}
\noindent
{\bf Acknowledgements}
\vskip .1in
\noindent
A.G. acknowledges Yoshio Koide for many helpful
discussions on possible texture of neutrino mass matrix
during his visit at University of
Shizuoka, Shizuoka, Japan.


\end{multicols}
\newpage
\begin{table}
\begin{tabular}{|c|c|}
{\rm Present Work}& {\rm Bahcall et.al \cite{Bahcall}}\\
\hline
$\Delta m^2_{12}$ & $\Delta m^{2_\odot}$ \\
(eV$^2$)          &      (eV$^2$)       \\
\hline
& \\ 
$1.39 \times 10^{-4}$ &
$5.0 \times 10^{-5}$ \\
& \\
\hline
\hline
$\Delta m^2_{23}$ & $\Delta m^2_{\rm atm}$ \\  
(eV$^2$)          &      (eV$^2$)       \\
\hline
& \\
$3.1 \times 10^{-3}$ & $3.1 \times 10^{-3}$ \\
& \\
\hline
\hline
$\tan^2\theta_{12}$ & $\tan^2\theta_\odot$ \\
\hline
& \\
0.42 & 0.42 \\
& \\
\end{tabular}
\caption{ 
Best fitted values for
neutrino oscillation parameters obtained
in the present work. The values estimated
in 
Ref.[1,5] 
are also given.
 }
\end{table}
\narrowtext
\begin{table}
\begin{center}
\begin{tabular}{|c|c|c|cc|}
\hline
Fields & $SU(2)_L\times U(1)_Y$ & $Z_2$ &
$Z_3$&\\
\hline
{\underline{\rm {leptons}}} &&&&\\
&&&&\\
$l_{1L}$& (2,-1)& 1 & $\omega$&\\
$l_{2L}$& (2,-1)& -1 & 1&\\
$l_{3L}$& (2,-1)& -1 & 1&\\
$e_{R}$& (1,-2)& 1 & $\omega$&\\
$\mu_{R}$& (1,-2)& 1 & 1&\\
$\tau_{R}$& (1,-2)& -1 & 1&\\
&&&&\\
{\underline{\rm {Scalars}}}&&&&\\
&&&&\\
$\phi_1$&(2,1)&1&1&\\
$\eta_1$&(1,0)&1& $\omega$&\\
$\eta_2$&(1,0)&-1& $\omega^2$&\\
&&&&\\
\hline
\end{tabular}
\end{center}
\begin{center}
\caption{
Representation content of the lepton and scalar
fields considered in the present model. The elements
of $Z_2$ and $Z_3$ are given by $\{1,-1\}$ , $\{1,
\omega, \omega^2 \}$ respectively. In general,
elements of $Z_n$ is given by
${\rm e}^{\frac{2\pi i}{\rm n}}$ .
}
\end{center}
\end{table}
\newpage
\begin{center}
{\bf Figure Caption}
\end{center}

\noindent Fig. 1 The region (shaded area) of the parameters 
$a$ and $b$ that produce the values of $\Delta m_{12}^2$ and 
$\tan^2\theta_{12}$ within 3$\sigma$ range of the best fitted 
values from global solar neutrino data analysis \cite{Bahcall}. 
The value for $\lambda$ is kept fixed at the best fit value in the 
present calculation. 
$\Delta m_{23}^2$ remains fixed at the best fit value for 
$\Delta m_{\rm atm}^2$. See text for details.  

\begin{thebibliography}{99}
\bibitem{Bahcall}
J.~N.~Bahcall, M.~C.~Gonzalez-Garcia, and C.
~Pe\~{n}a-Garay, eprint no. hep-ph/0204314.

\bibitem{sno} SNO Collaboration, Q.R.~Ahmad et al.,
nucl-ex/ 0204008, nucl-ex/0204009.

\bibitem{bimax}
V. Barger, S. Pakvasa, 
T. J. Weiller and  K. Whisnant, 
hep-ph/9806387, 
B. C. Allanach, hep-ph/9806294,
D. V. Ahluwalia, Mod. Phys. Lett. A {\bf 13}, 2249 (1998), 
I. Stancu and D. V. Ahluwalia, Phys. Lett. B{\bf 460}, 431 (1999),
 V. Barger, T. J. Weiller and K. Whisnant, 
hep-ph/9807319, J. Elwood, N. Irges and 
P. Ramond, Phys. Rev. Lett.{\bf 81}, 5064 (1998), 
hep-ph/9807228, E. Ma, 
Phys. Lett. B{\bf 442}, 238 (1998), hep-ph/
9807386, hep-ph/9902392, 
G. Alterelli and F. Feruglio, 
Phys. Lett. B {\bf439}, 112 (1998), hep-ph/9807353, 
Y. Nomura and T. Yanagida, Phys. Rev. 
D{\bf 59}, 017303 (1999), 
 hep-ph/
9807325,  A. S. Joshipura, hep-ph /9808261, 
A. S. Joshipura and S. Rindani, 
hep-ph/9811252, 
K. Oda et al., Phys. Rev. D{\bf 59}, 055001 (1999), 
hep-ph/9808241, H. Fritzsch and 
Z. Xing, hep-ph/9808272, J. Ellis 
et al., hep-ph/
9808301, A. S. Joshipura and S. Vempati, hep-ph /
 9808232, U. Sarkar, Phys. Rev. 
D{\bf 59}, 037302 (1999),
hep-ph/9808277,   
H. Georgi and S. Glashow , hep-ph/ 9808293;
A. Baltz, A. S. Goldhaber and M. Goldhaber, 
Phys. Rev. Lett. {\bf 81}, 5730 (1998), 
hep-ph/ 9806540, 
M. Jezabek and A. Sumino, 
Phys. Lett. B{\bf 440}, 327 (1998), hep-ph/ 9807310,  
S. Davidson and S. F. King, 
Phys. Lett. B{\bf 445}, 191 (1998),
hep-ph/ 9808296,
 K. Kang, S. K. Kang, C. S. Kim and  
 S. M. Kim, 
 hep-ph/ 9808419, S. Mohanty, 
D. P. Roy and  U. Sarkar, 
Phys. Lett. B{\bf 445}, 185 (1998), 
hep-ph/9808451;   
E. Ma, U. Sarkar and D. P. Roy, Phys. Lett. 
B{\bf 444}, 391 (1998), hep-ph/9810309,
B. Brahmachari, hep-ph/9808331,
R. N. Mohapatra and S. Nussinov, 
Phys. Lett B{\bf 441}, 299 (1998), 
hep-ph/ 9808301, Phys. Rev. 
D{\bf 60}, 031002 (1999), hep-ph/ 9809415,  
A. Ghosal, hep-ph/9903497, hep-ph/9905470, 
R. Barbieri, L.J.Hall and A. Strumia, 
Phys. Lett. B {\bf 445}, 407 (1999), 
hep-ph/9808333, Y. Grossmann, Y. Nir and 
Y. Shadmi, JHEP: {\bf 9810}, 007  (1998), 
hep-ph/9808355, 
C. Jarlskog, M. Matsuda, S. Skadhauge and M. 
Tanimoto, 
Phys. Lett. B{\bf 449}, 240 (1999), 
hep-ph/9812282, S. M. Bilenky and 
C. Giunti, hep-ph/9802201, 
C. Giunti, Phys. Rev. D{\bf 59}, 077301 (1999),  
hep-ph/ 
9810272, 
M. Fukugita, M. Tanimoto and T. Yanagida, 
Phys. Rev. D{\bf 57}, 4429 (1998), 
S. K. Kang and C. S. Kim, 
Phys. Rev. D {\bf 59}, 091302 (1999),  
R. N. Mohapatra, A. Perez-Lorenzana and C.A. de S. Pires, 
Phys. Lett. B{\bf 474}, 355 (2000), A. Aranda, 
C. D. Carone and R. F. Lebed , Phys. Lett. B{\bf 474}, 
170 (2000), hep-ph/0002044,
H. B. Benaoum and S. Nasri, hep-ph/9906232, 
C. H. Albright and S. M. Barr, hep-ph/9906297. 

\bibitem{ag1}
A. Ghosal, Phys. Rev.D {\bf 62}, 092001 (2000). 


\bibitem{SK}
Super-Kamiokande Collaboration, Y. Fukuda et al., Phys.
Lett. B {\bf 433}, 9 (1998), B {\bf 436}, 33 (1998).

\bibitem{oba}
Y.Obayashi, Neutrino Oscillations
and their origin, Fujiyoshida, Japan
( University Academy Press, Tokyo, 2000 ).

\bibitem{chooz}
M. Appollonio, et al., Phys. Lett.  B {\bf 420}, 397 (1998).

\bibitem{2beta} 
H. V. Klapdor-Kleingrothaus, A. Dietz, H. L. Harney,
and I. V. Krivosheina, Mod. Phys. Lett. A {\bf 16}
, 2409 (2002).

\bibitem{nonma}
H. V. Klapdor-Kleingrothaus, "60 Years of Double-Beta Decay, 
From Nuclear Physics to Beyond the Standard Model", 
World Scientific, Singapore (2001), 
R. N. Mohapatra and P. B. Pal, 'Massive Neutrinos in Physics 
and Astrophysics', World Scientific, Singapore, World Scientific 
lecture notes in Phyiscs, 41 (1991). 

\bibitem{reply}
see for references H. V. Klapdor-Kleingrothaus, hep-ph/0205228 .

\bibitem{two}
R. Barbieri, L. J. Hall
and A. Strumia, Phys. Lett. B{\bf 445}, 407 (1999),
 R. Barbieri et al.,
JHEP {\bf 9812} : 017 (1998),
C. D. Froggatt, M. Gibson and H. B. Nielsen,
Phys. Lett. B{\bf 446}, 256 (1999),
Y. Koide and
A. Ghosal, Phys. Rev. D {\bf 63}, 037301 (2001). 

\bibitem{yasue} T. Kitabayashi and M. Yasue, Phys.
Lett. {\bf B 524}, 308  (2002), hep-ph/0112287,
B. Brahmachari, S. Choubey and R. N. Mohapatra,
hep-ph/ 0204073, 
Y. Koide and A. Ghosal, hep-ph/0203113.

\bibitem{tex} 
E. K. Akhmedov,Phys. Lett. B{\bf 467}, 95 (1999), E. K. Akhmedov, G. 
C. Branco and M. N. Rebelo, Phys. Lett. B{\bf 478}, 215 (2000),
F. Feruglio, Nucl. Phys. Proc. Suppl. {\bf 100}, 299 (2001), 
G. Altarelli, F. Feruglio, hep-ph/0206077. 
\bibitem{ag} A. Ghosal,
Phys. Lett. {\bf B 398} (1997), 315.

\end{thebibliography}
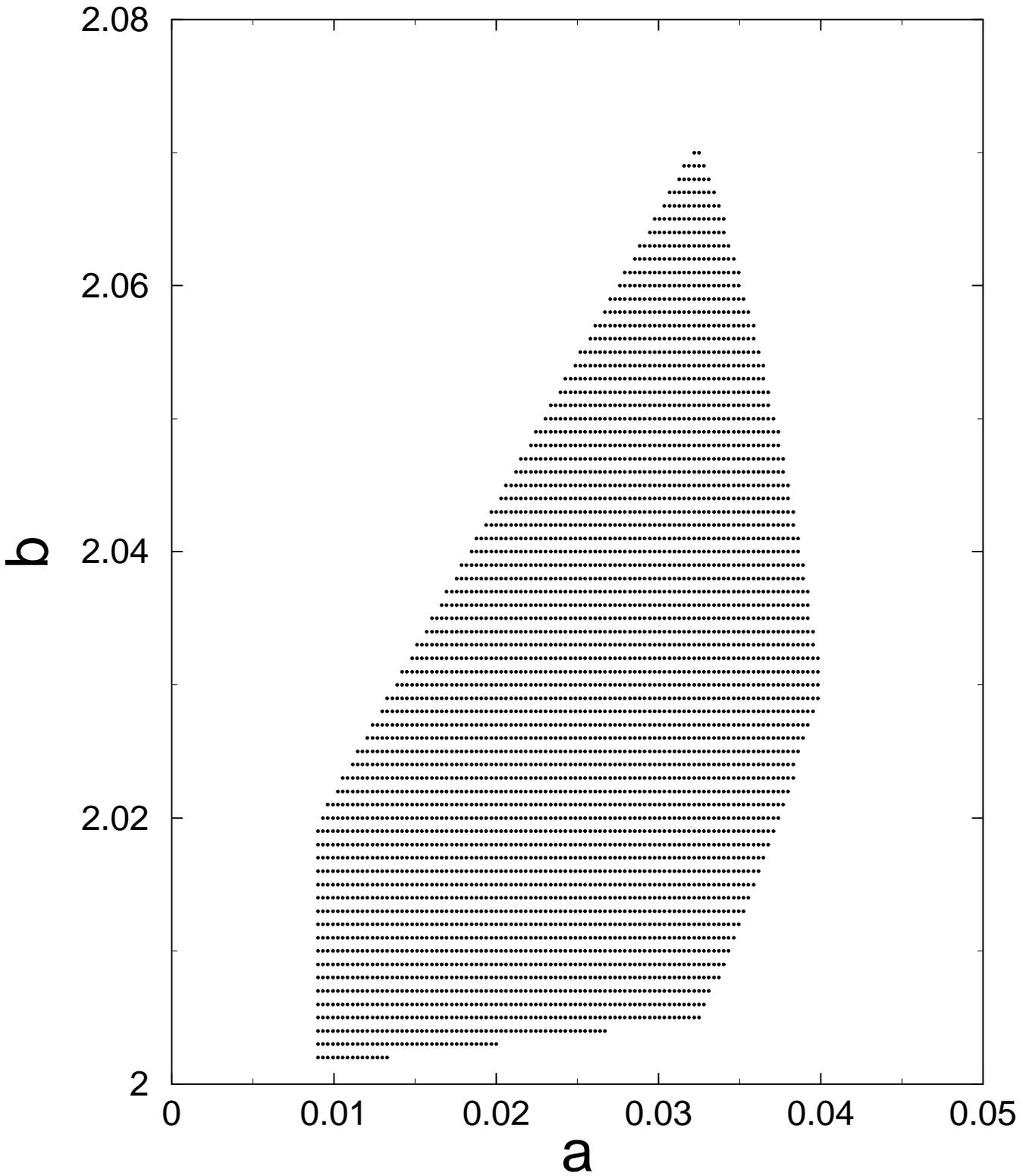
\end{document}